\documentclass[aps,prd,twocolumn,superscriptaddress,nofootinbib]{revtex4-1}

\usepackage{latexsym}
\usepackage{amsmath}
\usepackage{amssymb}
\usepackage{amsfonts}
\usepackage{bm}
\usepackage{physics}

\usepackage{color}
\definecolor{purple}{rgb}{0.5,0,0.5}
\definecolor{blue}{rgb}{0.0,0,0.9}
\definecolor{prdblue}{rgb}{0.133,0.118,0.498}
\usepackage[colorlinks=true, pdfstartview=FitV, linkcolor=prdblue, citecolor= prdblue, urlcolor=prdblue]{hyperref}

\usepackage{supertabular} 
\usepackage{placeins}
\usepackage{epsfig}
\usepackage{graphicx}

\begin{document}


\title{Diffusion Monte Carlo study of deuteron-like fully light hexaquarks}


\author{M.C. Gordillo}
\email[]{cgorbar@upo.es}
\affiliation{Departamento de Sistemas F\'isicos, Qu\'imicos y Naturales, Universidad Pablo de Olavide, Avenida Rectora Rosario Valpuesta 1, 41089, Dos Hermanas, Sevilla, Spain.}
\affiliation{Instituto Carlos I de Física Teórica y Computacional,
Universidad de Granada, E-18071 Granada, Spain.}

\author{J. Segovia}
\email[]{jsegovia@upo.es}
\affiliation{Departamento de Sistemas F\'isicos, Qu\'imicos y Naturales, Universidad Pablo de Olavide, Avenida Rectora Rosario Valpuesta 1, 41089, Dos Hermanas, Sevilla, Spain.}


\date{\today}

\begin{abstract}
We perform a Diffusion Monte Carlo study of fully light hexaquarks containing three $u$ quarks and three $d$ quarks within a constituent-quark model. Both compact and baryon--baryon-like arrangements were considered separately. All compact hexaquark configurations are found well  above their corresponding baryon--baryon thresholds, suggesting that deeply bound compact six-quark states are not favored in the light-quark sector within this model. By contrast, several dibaryon-like configurations lie close, but above, to the $NN$, $N\Delta$, and $\Delta\Delta$ thresholds and show spatial structures compatible with  molecular states. One configuration exhibits two well-defined nucleon-like subclusters separated by several femtometers, characteristic of deuteron-like spatial clustering, although its calculated energy remains slightly above the corresponding threshold.
\end{abstract}


\maketitle


The study of multiquark hadrons has experienced remarkable progress in recent years, driven by the increasing experimental evidence for exotic configurations beyond the conventional quark-model description of mesons and baryons~\cite{Gell-Mann:1964ewy, Zweig:1964CERN}. In particular, the observation of tetraquark- and pentaquark-like candidates by the Belle, BESIII, and LHCb collaborations has renewed interest in understanding the dynamics of strongly correlated multiquark systems  in terms of the underlying quark degrees of freedom~\cite{Dong:2020hxe, Chen:2016qju, Chen:2016spr, Guo:2017jvc, Liu:2019zoy, Yang:2020atz, Dong:2021bvy, Chen:2021erj, Cao:2023rhu, Mai:2022eur, Meng:2022ozq, Chen:2022asf, Guo:2022kdi, Ortega:2020tng, Huang:2023jec, Lebed:2023vnd, Zou:2021sha, Du:2021fmf, Liu:2024uxn, Johnson:2024omq, Entem:2025bqt, Hanhart:2025bun, Wang:2025dur, Wang:2025sic, Francis:2024fwf, Chen:2024eaq, Husken:2024rdk, Liu:2024uxn, Johnson:2024omq}. Despite these advances, the possible existence of stable or resonant hexaquark states remains an open question, especially in the sectors of light quarks.

The interest in six-quark states dates back to Jaffe's proposal of the H dibaryon~\cite{Jaffe:1976yi}, which triggered extensive experimental and theoretical activity in dibaryon spectroscopy~\cite{Mulders:1980vx, Kopeliovich:1990pp, Leandri:1995zm, Pepin:1998ih, Li:2000cb, Herzog:2003wt, Valcarce:2005em, Ikeda:2007nz, Olsen:2014qna, Francis:2018qch, Amarasinghe:2021lqa, Wan:2021vny, Wang:2024riu}. From the theoretical point of view, the description of six-quark systems requires non-perturbative techniques capable of accurately solving the corresponding many-body Schrodinger equation. Quantum Monte Carlo (QMC) methods have been successfully employed in several areas of physics, ranging from quantum chemistry and condensed matter to ultracold atomic gases and nuclear systems~\cite{Hammond:1994bk, Foulkes:2001zz, Nightingale:2014bk, Ceperley:1995zz, Giorgini:2008zz}.  However, their application to hadron physics has been comparatively limited, mainly because conventional hadrons are two- and three-body systems~\cite{Carlson:1982xi, Carlson:1983rw}. Nevertheless, the growing interest in multiquark spectroscopy has motivated the extension of these techniques to exotic hadrons containing four or more quarks~\cite{Bai:2016int, Gordillo:2020sgc}.

In this work, we investigate fully light hexaquark configurations within a constituent-quark model using the diffusion Monte Carlo (DMC) method. The DMC approach projects out the ground-state wave function through a stochastic evolution in imaginary time and provides, within statistical uncertainties, a numerically exact treatment of the correlated many-body problem defined by the adopted Hamiltonian and Hilbert-space truncation. This numerical accuracy should not be conflated with the phenomenological accuracy of the constituent-quark Hamiltonian or with that of the present restriction to total orbital angular momentum $L=0$. We analyze the masses and structural properties of six-quark systems containing three $u$ and three $d$ quarks, considering two different choices for the internal wave function. In the first one, all quarks are treated on the same footing, as appropriate for a compact six-quark configuration. In the second one, a baryon-baryon-like arrangement is introduced. This is not a mere product of colorless independent baryon functions, but includes hidden-color components in the internal wave function. Special attention is paid to the radial distribution functions of certain structures which are used to characterize the compact or molecular nature of the resulting six-quark states.

The non-relativistic six-body Hamiltonian that describes our six-quark system is:
\begin{equation}
H=\sum_{i=1}^{6}\left(m_i+\frac{{\bf p}_i^2}{2m_i}\right)-T_{\rm CM}+\sum_{i<j}V_{ij}({\bf r}_{ij}), 
\label{eq:Hamiltonian}
\end{equation}
where $m_i$ and ${\bf p}_i$ denote the mass and three-momentum of the $i$-th quark, respectively, $T_{\rm CM}$ is the center-of-mass kinetic-energy contribution, and ${\bf r}_{ij}={\bf r}_i-{\bf r}_j$ is the relative coordinate between quarks $i$ and $j$. The quark-quark interaction is described by the pairwise AL1 potential introduced in Refs.~\cite{Semay:1994ht, Silvestre-Brac:1996myf}, which contains the conventional Coulomb-like, linear confining, and smeared color-spin hyperfine terms and has been widely used in constituent-quark model calculations. Its parameters are fixed phenomenologically from conventional hadron spectroscopy and are not readjusted in the present six-quark calculation. As discussed in Ref.~\cite{Gordillo:2026usv}, the resulting nucleon mass is overestimated, giving $2M_N=2060\pm3$~MeV instead of the experimental threshold of approximately $1878$~MeV. In applications of the AL1 framework to baryons, an additional three-body term depending on the constituent-quark masses has been introduced phenomenologically to improve agreement with the experimental spectrum~\cite{Silvestre-Brac:1996myf}. We do not include this term because it is an \emph{ad hoc} correction to the pairwise Hamiltonian and no prescription has been given for extending it to multiquark systems beyond baryons~\cite{Gordillo:2026usv}. Consequently, the absolute energies reported below should not be interpreted as precision predictions for physical hexaquark masses. Their stability must instead be assessed relative to the corresponding two-baryon thresholds calculated consistently with the same Hamiltonian. We solve the Schr\"odinger equation associated with Eq.~\eqref{eq:Hamiltonian} using the DMC method, whose general implementation in multiquark systems is described in Ref.~\cite{Gordillo:2020sgc}.

The DMC algorithm requires the introduction of an approximation to the exact wave function of the system, usually referred to as the trial wave function. In the present approach, this wave function is constructed as a single state in the full Hilbert space,
\begin{align}
\Psi_{T}(\bm{R},\alpha)=\phi_T(\bm{R})\sum_{\alpha} d_{\alpha}\chi_{\alpha},
\label{eq:Eigenfunction}
\end{align}
where $\phi_T(\bm{R})$ depends on the quark coordinates and the functions $\chi_{\alpha}$ are direct products of eigenfunctions of the spin, isospin, and color operators,
\begin{align}
S^2 &= \left(\sum_{i=1}^{6} \frac{\vec{\sigma}_i}{2}\right)^2 , \nonumber \\
I^2 &= \left(\sum_{i=1}^{6} \frac{\vec{\lambda}_i^f}{2}\right)^2 , \nonumber \\
C^2 &= \left(\sum_{i=1}^{6} \frac{\vec{\lambda}_i^c}{2}\right)^2 ,
\label{eq:operators_penta}
\end{align}
with well-defined eigenvalues. Here, $\vec{\sigma}_i$ are the Pauli matrices acting on the spin of quark $i$, while $\vec{\lambda}^{c}_i$ are the standard Gell-Mann matrices acting in color space. The operators $\vec{\lambda}^{f}_i$ denote the generators of flavor SU(3). However, in the present work only the two-flavor $(u,d)$ sector is considered, so that only the components $\lambda_i^{f,1}$, $\lambda_i^{f,2}$, and $\lambda_i^{f,3}$, corresponding to the isospin SU(2) subgroup, contribute. For each set of spin, color, and isospin quantum numbers there can be several internal functions, whose contribution to the total internal wave function is included through the coefficients $d_{\alpha}$ in Eq.~\eqref{eq:Eigenfunction}. Further details of the procedure can be found in Ref.~\cite{Gordillo:2026usv}.

The Pauli principle is enforced by requiring antisymmetry under the exchange of identical quarks. Since the radial part is chosen to be totally symmetric, as discussed below, antisymmetry must be imposed on the spin-isospin-color wave function. A wave function is antisymmetric if it is an eigenvector with eigenvalue one of the operator
\begin{equation}
\mathcal{A} = \frac{1}{N_p} \sum_{\alpha=1}^{N_p} (-1)^P \mathcal{P}_{\alpha} , 
\label{eq:antisymope}
\end{equation}
where $N_p$ is the number of allowed permutations of the quark labels, $P$ is the parity of the permutation, and $\mathcal{P}_{\alpha}$ are the corresponding permutation matrices. Once the matrix representation of the operator in Eq.~\eqref{eq:antisymope} is constructed in the basis of spin-isospin-color wave functions defined above, its eigenvectors provide the allowed antisymmetric combinations within each sector. Different sets of antisymmetric functions are obtained depending on the choice of permutations included in Eq.~\eqref{eq:antisymope}. If all six quarks are treated on an equal footing, the resulting state corresponds to a compact hexaquark configuration. On the other hand, if only the first three quarks are antisymmetrized among themselves, and likewise the last three, while the two groups are kept distinguishable from each other, the result is a genuine color-singlet six-quark wave function with a dibaryon-like partition. It is not restricted to the baryon-singlet$\otimes$baryon-singlet channel, $[(qqq)_{\mathbf{1}_c}(qqq)_{\mathbf{1}_c}]_{\mathbf{1}_c}$, because its spin--isospin--color basis also contains configurations in which the individual three-quark subunits are not color singlets while the complete six-quark state is. These hidden-color components couple through the quark-quark interaction and provide correlations, and hence possible additional attraction, already in the $L=0$ sector that are absent from a pure product of two asymptotic color-singlet baryons. The functions associated with the two subunits are different and hence distinguishable, irrespective of whether they are orthogonal~\cite{TongTopicsQM} or not.

To complete the definition of the trial wave function in Eq.~\eqref{eq:Eigenfunction}, we specify the radial component $\phi_T(\bm{R})$. In this work, we restrict ourselves to states with total orbital angular momentum $L=0$. This restriction excludes $D$-wave components. In a conventional nucleon-level description of the physical deuteron, the tensor interaction couples the dominant ${}^{3}S_{1}$ component to the ${}^{3}D_{1}$ one, and the resulting $S$--$D$ mixing is important for a quantitative account of its binding and structure. The AL1 interaction employed here contains Coulomb-like, confining, and hyperfine spin-spin terms, but no tensor term capable of generating this physical ${}^{3}S_{1}$--${}^{3}D_{1}$ coupling. Accordingly, simply extending the spatial basis to $L=2$ without introducing the corresponding tensor dynamics would not provide a realistic description. Both ingredients would be required to resolve effects on the few-MeV scale characteristic of deuteron binding and could modify the precise location of the $(S,I)=(1,0)$ state relative to threshold. The present calculation should therefore be regarded as a study of deuteron-like spatial clustering within the $L=0$ six-quark sector, rather than as a complete dynamical description of the physical deuteron. The radial wave function depends only on the set of interparticle distances and is symmetric under permutations of the spatial coordinates. As in previous DMC studies of multiquark systems~\cite{Gordillo:2020sgc, Gordillo:2023tnz}, we use the form
\begin{equation}
\phi_T ({\bf R}) = \prod_{i<j} \exp(-a_{ij} r_{ij}) , 
\label{eq:radialwf}
\end{equation}
where the coefficients $a_{ij}$ are chosen to satisfy the cusp conditions associated with the Coulomb-like part of the potential for each pair of constituents.


\begin{table}[!t]
\caption{\label{tab:HexaSpectrum} Calculated energies, in MeV, for the lowest fully-light hexaquark configurations in the different spin--isospin channels considered in this work. Missing entries correspond to channels forbidden by symmetry constraints and ``$\cdots$" corresponds to DMC simulations in which the two baryons separate an infinite distance from each other. The quoted uncertainties represent the statistical errors of the DMC calculation. The last column provides a qualitative classification inferred from the closest two-baryon mass thresholds.}
\begin{ruledtabular}
\begin{tabular}{ccccc}
Configuration &  $(S,I)$ & $J^P$ & Energy (MeV) & State label~\cite{dyson} \\
\hline
Compact & $(1,0)$ & $1^+$ & $2716 \pm 7$ & $D_{01}$ \\
Compact & $(0,1)$ & $0^+$ & $2778 \pm 5$ & $D_{10}$ \\
Compact & $(3,0)$ & $3^+$ & $2878 \pm 6$ & $D_{03}$ \\
Compact & $(2,1)$ & $2^+$ & $2883 \pm 4$ & $D_{21}$ \\
Compact & $(1,2)$ & $1^+$ & $2992 \pm 2$ & $D_{21}$ \\
Compact & $(0,3)$ & $0^+$ & $3197 \pm 4$ & $D_{30}$ \\
\hline
\hline
Configuration & $(S,I)$ & $J^P$ & Energy (MeV) & Threshold\\
\hline
Di-baryon & $(0,0)$ & $0^+$ & $2078 \pm 6$ & $NN$-like \\
Di-baryon & $(0,1)$ & $0^+$ & $2086 \pm 9$ & $NN$-like \\
Di-baryon & $(0,2)$ & $0^+$ & $2764 \pm 2$ & -- \\
Di-baryon & $(0,3)$ & $0^+$ & $2625 \pm 5$ & $\Delta\Delta$-like \\
\hline
Di-baryon & $(1,0)$ & $1^+$ & $2072 \pm 8$ & $NN$-like \\
Di-baryon & $(1,1)$ & $1^+$ & $2091 \pm 4$ & $NN$-like \\
Di-baryon & $(1,2)$ & $1^+$ & $2347 \pm 5$ & $N\Delta$-like \\
Di-baryon & $(1,3)$ & $1^+$ & $2698 \pm 4$ & -- \\
\hline
Di-baryon & $(2,0)$ & $2^+$ & $2614 \pm 4$ & -- \\
Di-baryon & $(2,1)$ & $2^+$ & $2344 \pm 4$ & $N\Delta$-like \\
Di-baryon & $(2,2)$ & $2^+$ & $2352 \pm 5$ & $N\Delta$-like \\
Di-baryon & $(2,3)$ & $2^+$ & $\cdots$ & $\cdots$ \\
\hline
Di-baryon & $(3,0)$ & $3^+$ & $2857 \pm 4$ & -- \\
Di-baryon & $(3,1)$ & $3^+$ & $2864 \pm 5$ & -- \\
Di-baryon & $(3,2)$ & $3^+$ & $\cdots$ & $\cdots$ \\
Di-baryon & $(3,3)$ & $3^+$ & $\cdots$ & $\cdots$ \\
\end{tabular}
\end{ruledtabular}
\end{table}

We now discuss the spectrum obtained for the six-quark configurations in the $S=0,1,2,3$ and $I=0,1,2,3$ channels. The calculated masses are summarized in Table~\ref{tab:HexaSpectrum}. The quoted uncertainties correspond to the statistical errors associated with the DMC simulations.

In the fully antisymmetric quark sector, we consider only the configurations allowed by symmetry~\cite{dyson}. The lowest compact configuration is found in the $(S,I,J^P)=(1,0,1^+)$ channel, with a mass of $2716\pm7$ MeV. It is followed by the $(0,1,0^+)$ state, with a mass of $2778\pm5$ MeV. Both states lie above the nominal $\Delta\Delta$ threshold, which in the present model is $2M_\Delta = 2614$ MeV~\cite{Gordillo:2026usv}. At fixed isospin, the compact configurations listed in Table~\ref{tab:HexaSpectrum} exhibit a tendency for the higher-spin state to lie at higher energy: the $(3,0,3^+)$ and $(2,1,2^+)$ channels are located around $2.88$ GeV, above their $(1,0,1^+)$ and $(0,1,0^+)$ counterparts, respectively. This ordering should not, however, be interpreted as a universal relation between total spin and mass. Within the AL1 Hamiltonian, the explicit spin dependence originates from the color-spin hyperfine interaction, whose expectation value is determined not by $S$ alone but by the complete color-spin structure of the wave function, together with the isospin and Pauli constraints. For the compact $L=0$ states, these constraints restrict the allowed spin--isospin--color combinations and happen to produce the ordering shown in the table. Higher orbital angular momentum would alter the spatial permutation symmetry and, consequently, the allowed internal structures, so the same ordering need not persist. Indeed, previous DMC calculations found nearly degenerate $J^P=0^+$ and $1^+$ fully-heavy tetraquarks~\cite{Gordillo:2020sgc}, nearly degenerate isoscalar and isovector $J^{PC}=1^{++}$ $c\bar c n\bar n$ tetraquarks~\cite{Gordillo:2021bra}, and even a reversed spin ordering for compact fully-heavy pentaquarks~\cite{Gordillo:2024blx}. Hence, no general monotonic dependence $E=E(S)$ or $E=E(I)$ is implied by the model. The largest-isospin compact states, $(1,2,1^+)$ and $(0,3,0^+)$, lie around $3$ GeV or above. Therefore, none of these configurations are stable against dissociation into two-baryon channels within this model.

On the other hand, the dibaryon-like sector displays several configurations close to the corresponding two-baryon thresholds associated with the $NN$, $N\Delta$, and $\Delta\Delta$ channels. The available channels depend on the accessible values of the total spin and isospin of the two baryons into which the multiquark can split. If several such channels are available, we consider the lowest mass threshold. For instance, the $(S,I,J^P)=(0,0,0^+)$, $(1,0,1^+)$, $(0,1,0^+)$, and $(1,1,1^+)$ configurations lie moderately above the $2M_N = 2060$ MeV threshold~\cite{Gordillo:2026usv}. The apparent spin-dependent crossovers among these $NN$-like states are not statistically significant: at $I=0$, the $(0,0)$ and $(1,0)$ energies differ by only $6$ MeV, while at $I=1$, the $(0,1)$ and $(1,1)$ energies differ by only $5$ MeV. Both differences are comparable to the quoted DMC statistical uncertainties. More generally, the dibaryon energy reflects the interplay among the relevant two-baryon threshold, color-spin correlations, hidden-color components, and spatial clustering; no simple monotonic dependence on total spin is therefore expected. The $(0,3,0^+)$ configuration lies above, but close to, the $2M_\Delta$ threshold, whereas the $(1,2,1^+)$, $(2,1,2^+)$, and $(2,2,2^+)$ configurations dissociate into a nucleon and a $\Delta$ through a fall-apart process. The remaining combinations are either substantially higher in mass or separate infinitely and cannot be assigned to a particular two-baryon threshold. Overall, our results are qualitatively consistent with previous studies showing that deeply bound fully light hexaquark configurations are unstable with respect to the relevant baryon--baryon thresholds~\cite{Mulders:1980vx, Silvestre-Brac:1992xsl, Valcarce:2005em}.


\begin{figure}
\includegraphics[width=\linewidth, trim = 1.90cm 0.00cm 1.00cm 0.10cm, clip]{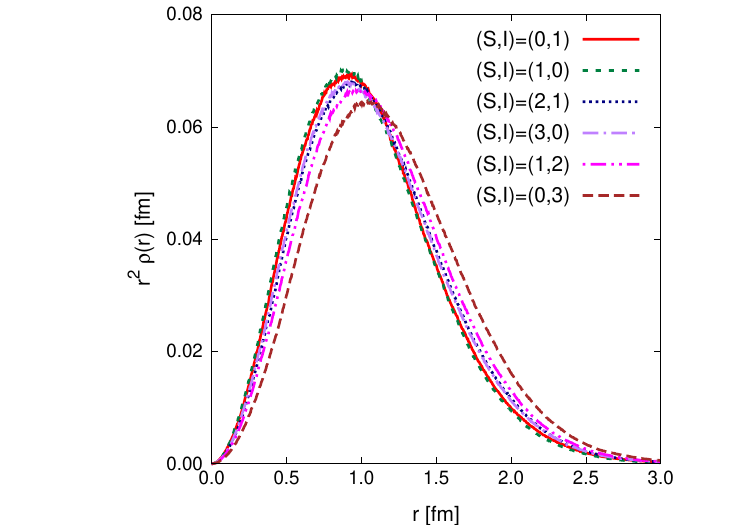}
\caption{\label{fig:CompactHexaquarks} Radial distribution functions $r^2\rho(r)$ for the compact fully-light hexaquark configurations studied in this work, where $r = |\bf{r}_2 - \bf{r}_1|$ denotes the relative distance between any two quarks. }
\end{figure}

\begin{figure*}
\includegraphics[width=0.45\textwidth, trim = 1.90cm 0.00cm 1.00cm 0.10cm, clip]{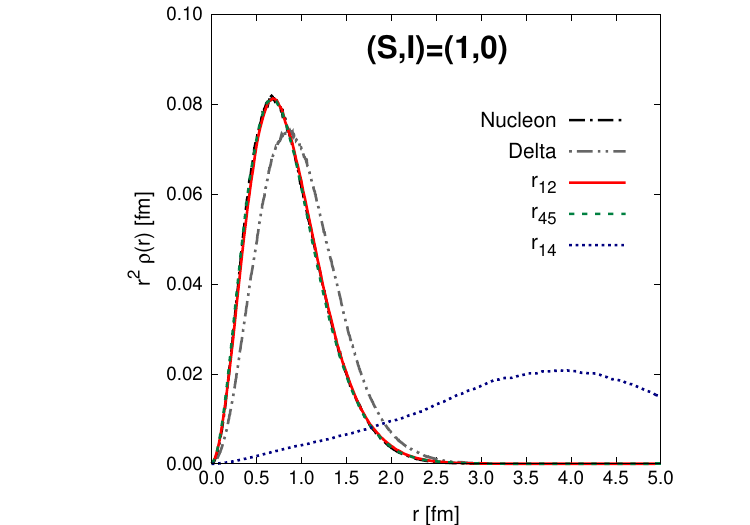}
\includegraphics[width=0.45\textwidth, trim = 1.90cm 0.00cm 1.00cm 0.10cm, clip]{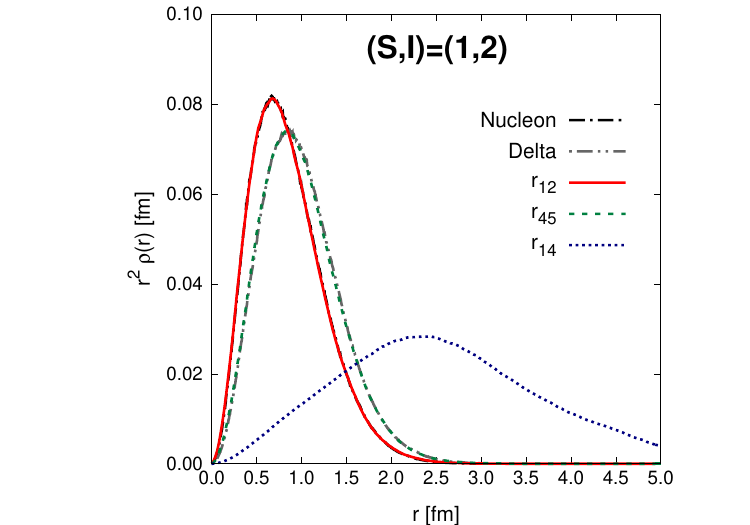} \\[1.0ex]
\includegraphics[width=0.45\textwidth, trim = 1.90cm 0.00cm 1.00cm 0.10cm, clip]{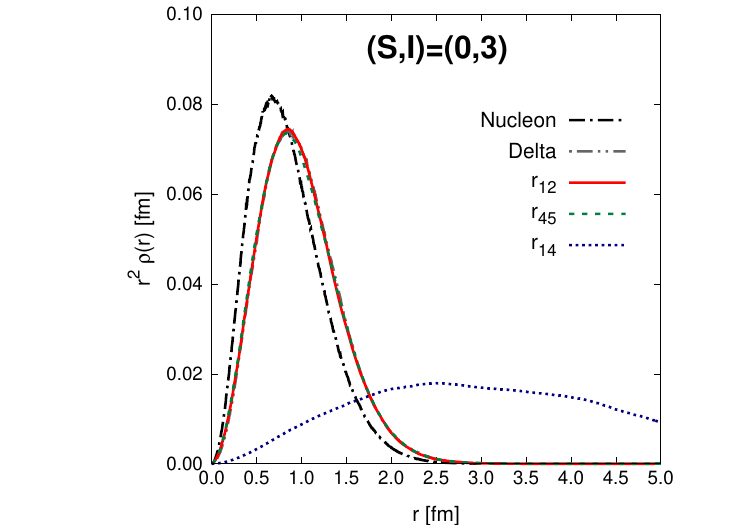} 
\includegraphics[width=0.45\textwidth, trim = 1.90cm 0.00cm 1.00cm 0.10cm, clip]{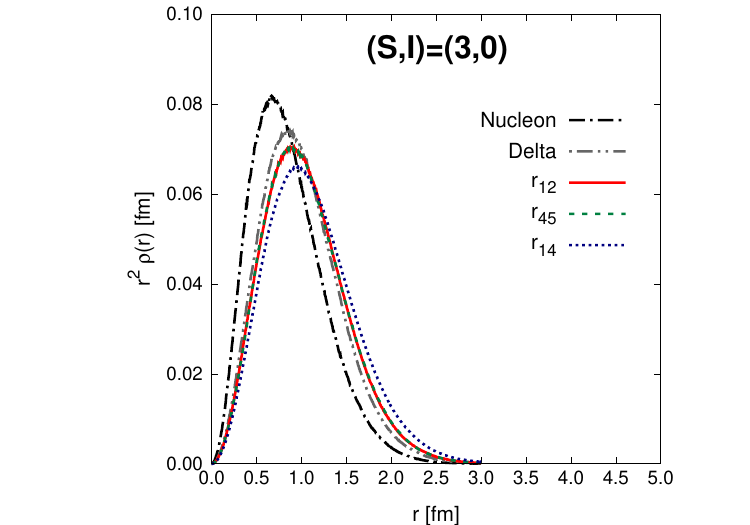} 

\caption{\label{fig:DiBaryons1} Same as Fig. \ref{fig:CompactHexaquarks} for different baryon-baryon-like configurations.}
\end{figure*}

There is a correlation between those mass results and the underlying structure of the multiquarks. Figure~\ref{fig:CompactHexaquarks} represents the probability of finding another quark at a given distance from a selected one in the fully antisymmetric structures. This magnitude is $r^2\rho(r)$, where $r=|{\bf r}_2-{\bf r}_1|$ denotes the relative distance between the two quarks and $\rho(r)$ is the quark density at a distance $r$. The fact that all quarks are equivalent in these compact structures means that we have a single curve for each spin--color--isospin configuration. All compact configurations exhibit localized distributions with maxima at distances of approximately $1~\text{fm}$. Although the broader profiles in Fig.~\ref{fig:CompactHexaquarks} generally correspond to configurations with larger $S$ and/or $I$, both quantum numbers and the underlying color-spin structures change simultaneously. The relatively small variations should therefore not be interpreted as a monotonic dependence on either $S$ or $I$ alone.

In Fig.~\ref{fig:DiBaryons1} we show the most representative examples of baryon--baryon-like configurations. Since the system contains two sets of indistinguishable quarks, there are, a priori, three distinct characteristic interquark distances. The distance $r_{12}$ characterizes correlations between quarks belonging to the first baryonic subunit, whereas $r_{45}$ describes correlations within the second one. Finally, $r_{14}$ measures correlations between quarks belonging to different subunits and may therefore be used as a proxy for the separation between the centers of mass of the two baryonic clusters, although the two quantities are not strictly equivalent.  For the sake of comparison, we display also the distributions corresponding to a nucleon and a $\Delta$ baryon. 

What we see is that the structures correlate perfectly to the findings in the mass sector. For instance, in the $(1,0,1^+)$ case,  the two baryonic structures correspond to two nucleons separated by a distance whose maximum is $\sim$ 4 fm,  in perfect agreement with its $NN$ mass threshold, according to Table~\ref{tab:HexaSpectrum}.  The $(1,2,1^+)$ hexaquark have one of the units similar to a nucleon, and another as a $\Delta$, while the $(0,3,0^+)$ structure corresponds to two $\Delta$'s joined together.  The final hexaquark,  $(3,0,1^+)$ is a compact object that cannot be separated into two subunits and whose mass is similar to the analogous $D_{03}$ state.  At this point we have to stress that the fact that those subunits remain close to each other is not an artifact of the form of the spatial wavefunction used (Eq. \ref{eq:radialwf}), since the DMC algorithm is able to split the system into two when both pieces are colorless~\cite{Gordillo:2024blx}. 


Summarizing, we have studied both the masses and configurations of all possible $uuuddd$ hexaquark states using the DMC technique. We find that the masses and structures are strongly correlated, with several near-threshold configurations. The $(S,I,J^P)=(1,0,1^+)$ state is particularly interesting: its radial distributions reveal two well-defined nucleon-like subclusters separated by a comparatively large distance. Here, the term ``deuteron-like'' refers specifically to this emergent two-nucleon spatial clustering in a genuine six-quark $L=0$ wave function, whose hidden-color components can provide correlations beyond a pure $NN\;{}^3S_1$ product state. It does not imply a complete dynamical description of the physical deuteron. The calculated energy, $2072\pm8$ MeV, lies $12\pm9$ MeV above the consistently calculated model threshold, $2M_N=2060\pm3$ MeV. By comparison, the physical deuteron has a binding energy $B_d\simeq2.2246$ MeV~\cite{Kessler1999}. The present $L=0$ calculation, with an interaction that lacks tensor-induced ${}^{3}S_1$--${}^{3}D_1$ mixing, cannot resolve dynamics reliably on this few-MeV scale. Including both higher partial waves and the corresponding tensor interaction may shift the precise position of the state relative to threshold; therefore, the conclusion that it is slightly unbound applies only within the adopted $L=0$ AL1 framework.

\begin{acknowledgments}
We acknowledge financial support from 
Ministerio Espa\~nol de Ciencia e Innovaci\'on under grant Nos. PID2020-113565GB-C22 and PID2022-140440NB-C22;
Junta de Andaluc\'ia under contract Nos. PAIDI FQM-205 and FQM-370.
The authors acknowledge, too, the use of the computer facilities of C3UPO at the Universidad Pablo de Olavide (UPO), de Sevilla within ``Plan Propio de Investigación y Transferencia (2023-2026) de la UPO, por la Consejería de Universidades, Investigación e Innovación de la Junta de Andalucía y por el Programa FEDER Andalucía 2021-2027, 2023/00002/014".
\end{acknowledgments}


\bibliography{print_DMC-FullyLightHexaquarks_vShort}

\begin{thebibliography}{62}%
\makeatletter
\providecommand \@ifxundefined [1]{%
 \@ifx{#1\undefined}
}%
\providecommand \@ifnum [1]{%
 \ifnum #1\expandafter \@firstoftwo
 \else \expandafter \@secondoftwo
 \fi
}%
\providecommand \@ifx [1]{%
 \ifx #1\expandafter \@firstoftwo
 \else \expandafter \@secondoftwo
 \fi
}%
\providecommand \natexlab [1]{#1}%
\providecommand \enquote  [1]{``#1''}%
\providecommand \bibnamefont  [1]{#1}%
\providecommand \bibfnamefont [1]{#1}%
\providecommand \citenamefont [1]{#1}%
\providecommand \href@noop [0]{\@secondoftwo}%
\providecommand \href [0]{\begingroup \@sanitize@url \@href}%
\providecommand \@href[1]{\@@startlink{#1}\@@href}%
\providecommand \@@href[1]{\endgroup#1\@@endlink}%
\providecommand \@sanitize@url [0]{\catcode `\\12\catcode `\$12\catcode
  `\&12\catcode `\#12\catcode `\^12\catcode `\_12\catcode `\%12\relax}%
\providecommand \@@startlink[1]{}%
\providecommand \@@endlink[0]{}%
\providecommand \url  [0]{\begingroup\@sanitize@url \@url }%
\providecommand \@url [1]{\endgroup\@href {#1}{\urlprefix }}%
\providecommand \urlprefix  [0]{URL }%
\providecommand \Eprint [0]{\href }%
\providecommand \doibase [0]{http://dx.doi.org/}%
\providecommand \selectlanguage [0]{\@gobble}%
\providecommand \bibinfo  [0]{\@secondoftwo}%
\providecommand \bibfield  [0]{\@secondoftwo}%
\providecommand \translation [1]{[#1]}%
\providecommand \BibitemOpen [0]{}%
\providecommand \bibitemStop [0]{}%
\providecommand \bibitemNoStop [0]{.\EOS\space}%
\providecommand \EOS [0]{\spacefactor3000\relax}%
\providecommand \BibitemShut  [1]{\csname bibitem#1\endcsname}%
\let\auto@bib@innerbib\@empty
\bibitem [{\citenamefont {Gell-Mann}(1964)}]{Gell-Mann:1964ewy}%
  \BibitemOpen
  \bibfield  {author} {\bibinfo {author} {\bibfnamefont {M.}~\bibnamefont
  {Gell-Mann}},\ }\href {\doibase 10.1016/S0031-9163(64)92001-3} {\bibfield
  {journal} {\bibinfo  {journal} {Phys. Lett.}\ }\textbf {\bibinfo {volume}
  {8}},\ \bibinfo {pages} {214} (\bibinfo {year} {1964})}\BibitemShut {NoStop}%
\bibitem [{\citenamefont {Zweig}(1964)}]{Zweig:1964CERN}%
  \BibitemOpen
  \bibfield  {author} {\bibinfo {author} {\bibfnamefont {G.}~\bibnamefont
  {Zweig}},\ }\href@noop {} {\bibfield  {journal} {\bibinfo  {journal} {CERN
  Report No.8182/TH.401, CERN Report No.8419/TH.412}\ } (\bibinfo {year}
  {1964})}\BibitemShut {NoStop}%
\bibitem [{\citenamefont {Dong}\ \emph
  {et~al.}(2021{\natexlab{a}})\citenamefont {Dong}, \citenamefont {Guo},\ and\
  \citenamefont {Zou}}]{Dong:2020hxe}%
  \BibitemOpen
  \bibfield  {author} {\bibinfo {author} {\bibfnamefont {X.-K.}\ \bibnamefont
  {Dong}}, \bibinfo {author} {\bibfnamefont {F.-K.}\ \bibnamefont {Guo}}, \
  and\ \bibinfo {author} {\bibfnamefont {B.-S.}\ \bibnamefont {Zou}},\ }\href
  {\doibase 10.1103/PhysRevLett.126.152001} {\bibfield  {journal} {\bibinfo
  {journal} {Phys. Rev. Lett.}\ }\textbf {\bibinfo {volume} {126}},\ \bibinfo
  {pages} {152001} (\bibinfo {year} {2021}{\natexlab{a}})},\ \Eprint
  {http://arxiv.org/abs/2011.14517} {arXiv:2011.14517 [hep-ph]} \BibitemShut
  {NoStop}%
\bibitem [{\citenamefont {Chen}\ \emph {et~al.}(2016)\citenamefont {Chen},
  \citenamefont {Chen}, \citenamefont {Liu},\ and\ \citenamefont
  {Zhu}}]{Chen:2016qju}%
  \BibitemOpen
  \bibfield  {author} {\bibinfo {author} {\bibfnamefont {H.-X.}\ \bibnamefont
  {Chen}}, \bibinfo {author} {\bibfnamefont {W.}~\bibnamefont {Chen}}, \bibinfo
  {author} {\bibfnamefont {X.}~\bibnamefont {Liu}}, \ and\ \bibinfo {author}
  {\bibfnamefont {S.-L.}\ \bibnamefont {Zhu}},\ }\href {\doibase
  10.1016/j.physrep.2016.05.004} {\bibfield  {journal} {\bibinfo  {journal}
  {Phys. Rept.}\ }\textbf {\bibinfo {volume} {639}},\ \bibinfo {pages} {1}
  (\bibinfo {year} {2016})},\ \Eprint {http://arxiv.org/abs/1601.02092}
  {arXiv:1601.02092 [hep-ph]} \BibitemShut {NoStop}%
\bibitem [{\citenamefont {Chen}\ \emph {et~al.}(2017)\citenamefont {Chen},
  \citenamefont {Chen}, \citenamefont {Liu}, \citenamefont {Liu},\ and\
  \citenamefont {Zhu}}]{Chen:2016spr}%
  \BibitemOpen
  \bibfield  {author} {\bibinfo {author} {\bibfnamefont {H.-X.}\ \bibnamefont
  {Chen}}, \bibinfo {author} {\bibfnamefont {W.}~\bibnamefont {Chen}}, \bibinfo
  {author} {\bibfnamefont {X.}~\bibnamefont {Liu}}, \bibinfo {author}
  {\bibfnamefont {Y.-R.}\ \bibnamefont {Liu}}, \ and\ \bibinfo {author}
  {\bibfnamefont {S.-L.}\ \bibnamefont {Zhu}},\ }\href {\doibase
  10.1088/1361-6633/aa6420} {\bibfield  {journal} {\bibinfo  {journal} {Rept.
  Prog. Phys.}\ }\textbf {\bibinfo {volume} {80}},\ \bibinfo {pages} {076201}
  (\bibinfo {year} {2017})},\ \Eprint {http://arxiv.org/abs/1609.08928}
  {arXiv:1609.08928 [hep-ph]} \BibitemShut {NoStop}%
\bibitem [{\citenamefont {Guo}\ \emph {et~al.}(2018)\citenamefont {Guo},
  \citenamefont {Hanhart}, \citenamefont {Mei\ss{}ner}, \citenamefont {Wang},
  \citenamefont {Zhao},\ and\ \citenamefont {Zou}}]{Guo:2017jvc}%
  \BibitemOpen
  \bibfield  {author} {\bibinfo {author} {\bibfnamefont {F.-K.}\ \bibnamefont
  {Guo}}, \bibinfo {author} {\bibfnamefont {C.}~\bibnamefont {Hanhart}},
  \bibinfo {author} {\bibfnamefont {U.-G.}\ \bibnamefont {Mei\ss{}ner}},
  \bibinfo {author} {\bibfnamefont {Q.}~\bibnamefont {Wang}}, \bibinfo {author}
  {\bibfnamefont {Q.}~\bibnamefont {Zhao}}, \ and\ \bibinfo {author}
  {\bibfnamefont {B.-S.}\ \bibnamefont {Zou}},\ }\href {\doibase
  10.1103/RevModPhys.90.015004} {\bibfield  {journal} {\bibinfo  {journal}
  {Rev. Mod. Phys.}\ }\textbf {\bibinfo {volume} {90}},\ \bibinfo {pages}
  {015004} (\bibinfo {year} {2018})},\ \Eprint
  {http://arxiv.org/abs/1705.00141} {arXiv:1705.00141 [hep-ph]} \BibitemShut
  {NoStop}%
\bibitem [{\citenamefont {Liu}\ \emph {et~al.}(2019)\citenamefont {Liu},
  \citenamefont {Chen}, \citenamefont {Chen}, \citenamefont {Liu},\ and\
  \citenamefont {Zhu}}]{Liu:2019zoy}%
  \BibitemOpen
  \bibfield  {author} {\bibinfo {author} {\bibfnamefont {Y.-R.}\ \bibnamefont
  {Liu}}, \bibinfo {author} {\bibfnamefont {H.-X.}\ \bibnamefont {Chen}},
  \bibinfo {author} {\bibfnamefont {W.}~\bibnamefont {Chen}}, \bibinfo {author}
  {\bibfnamefont {X.}~\bibnamefont {Liu}}, \ and\ \bibinfo {author}
  {\bibfnamefont {S.-L.}\ \bibnamefont {Zhu}},\ }\href {\doibase
  10.1016/j.ppnp.2019.04.003} {\bibfield  {journal} {\bibinfo  {journal} {Prog.
  Part. Nucl. Phys.}\ }\textbf {\bibinfo {volume} {107}},\ \bibinfo {pages}
  {237} (\bibinfo {year} {2019})},\ \Eprint {http://arxiv.org/abs/1903.11976}
  {arXiv:1903.11976 [hep-ph]} \BibitemShut {NoStop}%
\bibitem [{\citenamefont {Yang}\ \emph {et~al.}(2020)\citenamefont {Yang},
  \citenamefont {Ping},\ and\ \citenamefont {Segovia}}]{Yang:2020atz}%
  \BibitemOpen
  \bibfield  {author} {\bibinfo {author} {\bibfnamefont {G.}~\bibnamefont
  {Yang}}, \bibinfo {author} {\bibfnamefont {J.}~\bibnamefont {Ping}}, \ and\
  \bibinfo {author} {\bibfnamefont {J.}~\bibnamefont {Segovia}},\ }\href
  {\doibase 10.3390/sym12111869} {\bibfield  {journal} {\bibinfo  {journal}
  {Symmetry}\ }\textbf {\bibinfo {volume} {12}},\ \bibinfo {pages} {1869}
  (\bibinfo {year} {2020})},\ \Eprint {http://arxiv.org/abs/2009.00238}
  {arXiv:2009.00238 [hep-ph]} \BibitemShut {NoStop}%
\bibitem [{\citenamefont {Dong}\ \emph
  {et~al.}(2021{\natexlab{b}})\citenamefont {Dong}, \citenamefont {Guo},\ and\
  \citenamefont {Zou}}]{Dong:2021bvy}%
  \BibitemOpen
  \bibfield  {author} {\bibinfo {author} {\bibfnamefont {X.-K.}\ \bibnamefont
  {Dong}}, \bibinfo {author} {\bibfnamefont {F.-K.}\ \bibnamefont {Guo}}, \
  and\ \bibinfo {author} {\bibfnamefont {B.-S.}\ \bibnamefont {Zou}},\ }\href
  {\doibase 10.1088/1572-9494/ac27a2} {\bibfield  {journal} {\bibinfo
  {journal} {Commun. Theor. Phys.}\ }\textbf {\bibinfo {volume} {73}},\
  \bibinfo {pages} {125201} (\bibinfo {year} {2021}{\natexlab{b}})},\ \Eprint
  {http://arxiv.org/abs/2108.02673} {arXiv:2108.02673 [hep-ph]} \BibitemShut
  {NoStop}%
\bibitem [{\citenamefont {Chen}(2022)}]{Chen:2021erj}%
  \BibitemOpen
  \bibfield  {author} {\bibinfo {author} {\bibfnamefont {H.-X.}\ \bibnamefont
  {Chen}},\ }\href {\doibase 10.1103/PhysRevD.105.094003} {\bibfield  {journal}
  {\bibinfo  {journal} {Phys. Rev. D}\ }\textbf {\bibinfo {volume} {105}},\
  \bibinfo {pages} {094003} (\bibinfo {year} {2022})},\ \Eprint
  {http://arxiv.org/abs/2103.08586} {arXiv:2103.08586 [hep-ph]} \BibitemShut
  {NoStop}%
\bibitem [{\citenamefont {Cao}(2023)}]{Cao:2023rhu}%
  \BibitemOpen
  \bibfield  {author} {\bibinfo {author} {\bibfnamefont {X.}~\bibnamefont
  {Cao}},\ }\href@noop {} {\  (\bibinfo {year} {2023})},\ \Eprint
  {http://arxiv.org/abs/2301.11253} {arXiv:2301.11253 [hep-ph]} \BibitemShut
  {NoStop}%
\bibitem [{\citenamefont {Mai}\ \emph {et~al.}(2023)\citenamefont {Mai},
  \citenamefont {Mei\ss{}ner},\ and\ \citenamefont {Urbach}}]{Mai:2022eur}%
  \BibitemOpen
  \bibfield  {author} {\bibinfo {author} {\bibfnamefont {M.}~\bibnamefont
  {Mai}}, \bibinfo {author} {\bibfnamefont {U.-G.}\ \bibnamefont
  {Mei\ss{}ner}}, \ and\ \bibinfo {author} {\bibfnamefont {C.}~\bibnamefont
  {Urbach}},\ }\href {\doibase 10.1016/j.physrep.2022.11.005} {\bibfield
  {journal} {\bibinfo  {journal} {Phys. Rept.}\ }\textbf {\bibinfo {volume}
  {1001}},\ \bibinfo {pages} {1} (\bibinfo {year} {2023})},\ \Eprint
  {http://arxiv.org/abs/2206.01477} {arXiv:2206.01477 [hep-ph]} \BibitemShut
  {NoStop}%
\bibitem [{\citenamefont {Meng}\ \emph {et~al.}(2022)\citenamefont {Meng},
  \citenamefont {Wang}, \citenamefont {Wang},\ and\ \citenamefont
  {Zhu}}]{Meng:2022ozq}%
  \BibitemOpen
  \bibfield  {author} {\bibinfo {author} {\bibfnamefont {L.}~\bibnamefont
  {Meng}}, \bibinfo {author} {\bibfnamefont {B.}~\bibnamefont {Wang}}, \bibinfo
  {author} {\bibfnamefont {G.-J.}\ \bibnamefont {Wang}}, \ and\ \bibinfo
  {author} {\bibfnamefont {S.-L.}\ \bibnamefont {Zhu}},\ }\href@noop {} {\
  (\bibinfo {year} {2022})},\ \Eprint {http://arxiv.org/abs/2204.08716}
  {arXiv:2204.08716 [hep-ph]} \BibitemShut {NoStop}%
\bibitem [{\citenamefont {Chen}\ \emph {et~al.}(2023)\citenamefont {Chen},
  \citenamefont {Chen}, \citenamefont {Liu}, \citenamefont {Liu},\ and\
  \citenamefont {Zhu}}]{Chen:2022asf}%
  \BibitemOpen
  \bibfield  {author} {\bibinfo {author} {\bibfnamefont {H.-X.}\ \bibnamefont
  {Chen}}, \bibinfo {author} {\bibfnamefont {W.}~\bibnamefont {Chen}}, \bibinfo
  {author} {\bibfnamefont {X.}~\bibnamefont {Liu}}, \bibinfo {author}
  {\bibfnamefont {Y.-R.}\ \bibnamefont {Liu}}, \ and\ \bibinfo {author}
  {\bibfnamefont {S.-L.}\ \bibnamefont {Zhu}},\ }\href {\doibase
  10.1088/1361-6633/aca3b6} {\bibfield  {journal} {\bibinfo  {journal} {Rept.
  Prog. Phys.}\ }\textbf {\bibinfo {volume} {86}},\ \bibinfo {pages} {026201}
  (\bibinfo {year} {2023})},\ \Eprint {http://arxiv.org/abs/2204.02649}
  {arXiv:2204.02649 [hep-ph]} \BibitemShut {NoStop}%
\bibitem [{\citenamefont {Guo}\ \emph {et~al.}(2022)\citenamefont {Guo},
  \citenamefont {Peng}, \citenamefont {Xie},\ and\ \citenamefont
  {Zhou}}]{Guo:2022kdi}%
  \BibitemOpen
  \bibfield  {author} {\bibinfo {author} {\bibfnamefont {F.-K.}\ \bibnamefont
  {Guo}}, \bibinfo {author} {\bibfnamefont {H.}~\bibnamefont {Peng}}, \bibinfo
  {author} {\bibfnamefont {J.-J.}\ \bibnamefont {Xie}}, \ and\ \bibinfo
  {author} {\bibfnamefont {X.}~\bibnamefont {Zhou}},\ }\href@noop {} {\
  (\bibinfo {year} {2022})},\ \Eprint {http://arxiv.org/abs/2203.07141}
  {arXiv:2203.07141 [hep-ph]} \BibitemShut {NoStop}%
\bibitem [{\citenamefont {Ortega}\ and\ \citenamefont
  {Entem}(2021)}]{Ortega:2020tng}%
  \BibitemOpen
  \bibfield  {author} {\bibinfo {author} {\bibfnamefont {P.~G.}\ \bibnamefont
  {Ortega}}\ and\ \bibinfo {author} {\bibfnamefont {D.~R.}\ \bibnamefont
  {Entem}},\ }\href {\doibase 10.3390/sym13020279} {\bibfield  {journal}
  {\bibinfo  {journal} {Symmetry}\ }\textbf {\bibinfo {volume} {13}},\ \bibinfo
  {pages} {279} (\bibinfo {year} {2021})},\ \Eprint
  {http://arxiv.org/abs/2012.10105} {arXiv:2012.10105 [hep-ph]} \BibitemShut
  {NoStop}%
\bibitem [{\citenamefont {Huang}\ \emph {et~al.}(2023)\citenamefont {Huang},
  \citenamefont {Deng}, \citenamefont {Liu}, \citenamefont {Tan},\ and\
  \citenamefont {Ping}}]{Huang:2023jec}%
  \BibitemOpen
  \bibfield  {author} {\bibinfo {author} {\bibfnamefont {H.}~\bibnamefont
  {Huang}}, \bibinfo {author} {\bibfnamefont {C.}~\bibnamefont {Deng}},
  \bibinfo {author} {\bibfnamefont {X.}~\bibnamefont {Liu}}, \bibinfo {author}
  {\bibfnamefont {Y.}~\bibnamefont {Tan}}, \ and\ \bibinfo {author}
  {\bibfnamefont {J.}~\bibnamefont {Ping}},\ }\href {\doibase
  10.3390/sym15071298} {\bibfield  {journal} {\bibinfo  {journal} {Symmetry}\
  }\textbf {\bibinfo {volume} {15}},\ \bibinfo {pages} {1298} (\bibinfo {year}
  {2023})}\BibitemShut {NoStop}%
\bibitem [{\citenamefont {Lebed}(2023)}]{Lebed:2023vnd}%
  \BibitemOpen
  \bibfield  {author} {\bibinfo {author} {\bibfnamefont {R.~F.}\ \bibnamefont
  {Lebed}},\ }\href {\doibase 10.22323/1.445.0028} {\bibfield  {journal}
  {\bibinfo  {journal} {PoS}\ }\textbf {\bibinfo {volume} {FPCP2023}},\
  \bibinfo {pages} {028} (\bibinfo {year} {2023})},\ \Eprint
  {http://arxiv.org/abs/2308.00781} {arXiv:2308.00781 [hep-ph]} \BibitemShut
  {NoStop}%
\bibitem [{\citenamefont {Zou}(2021)}]{Zou:2021sha}%
  \BibitemOpen
  \bibfield  {author} {\bibinfo {author} {\bibfnamefont {B.-S.}\ \bibnamefont
  {Zou}},\ }\href {\doibase 10.1016/j.scib.2021.04.023} {\bibfield  {journal}
  {\bibinfo  {journal} {Sci. Bull.}\ }\textbf {\bibinfo {volume} {66}},\
  \bibinfo {pages} {1258} (\bibinfo {year} {2021})},\ \Eprint
  {http://arxiv.org/abs/2103.15273} {arXiv:2103.15273 [hep-ph]} \BibitemShut
  {NoStop}%
\bibitem [{\citenamefont {Du}\ \emph {et~al.}(2021)\citenamefont {Du},
  \citenamefont {Baru}, \citenamefont {Guo}, \citenamefont {Hanhart},
  \citenamefont {Mei\ss{}ner}, \citenamefont {Oller},\ and\ \citenamefont
  {Wang}}]{Du:2021fmf}%
  \BibitemOpen
  \bibfield  {author} {\bibinfo {author} {\bibfnamefont {M.-L.}\ \bibnamefont
  {Du}}, \bibinfo {author} {\bibfnamefont {V.}~\bibnamefont {Baru}}, \bibinfo
  {author} {\bibfnamefont {F.-K.}\ \bibnamefont {Guo}}, \bibinfo {author}
  {\bibfnamefont {C.}~\bibnamefont {Hanhart}}, \bibinfo {author} {\bibfnamefont
  {U.-G.}\ \bibnamefont {Mei\ss{}ner}}, \bibinfo {author} {\bibfnamefont
  {J.~A.}\ \bibnamefont {Oller}}, \ and\ \bibinfo {author} {\bibfnamefont
  {Q.}~\bibnamefont {Wang}},\ }\href {\doibase 10.1007/JHEP08(2021)157}
  {\bibfield  {journal} {\bibinfo  {journal} {JHEP}\ }\textbf {\bibinfo
  {volume} {08}},\ \bibinfo {pages} {157} (\bibinfo {year} {2021})},\ \Eprint
  {http://arxiv.org/abs/2102.07159} {arXiv:2102.07159 [hep-ph]} \BibitemShut
  {NoStop}%
\bibitem [{\citenamefont {Liu}\ \emph {et~al.}(2024)\citenamefont {Liu},
  \citenamefont {Pan}, \citenamefont {Liu}, \citenamefont {Wu}, \citenamefont
  {Lu},\ and\ \citenamefont {Geng}}]{Liu:2024uxn}%
  \BibitemOpen
  \bibfield  {author} {\bibinfo {author} {\bibfnamefont {M.-Z.}\ \bibnamefont
  {Liu}}, \bibinfo {author} {\bibfnamefont {Y.-W.}\ \bibnamefont {Pan}},
  \bibinfo {author} {\bibfnamefont {Z.-W.}\ \bibnamefont {Liu}}, \bibinfo
  {author} {\bibfnamefont {T.-W.}\ \bibnamefont {Wu}}, \bibinfo {author}
  {\bibfnamefont {J.-X.}\ \bibnamefont {Lu}}, \ and\ \bibinfo {author}
  {\bibfnamefont {L.-S.}\ \bibnamefont {Geng}},\ }\href@noop {} {\  (\bibinfo
  {year} {2024})},\ \Eprint {http://arxiv.org/abs/2404.06399} {arXiv:2404.06399
  [hep-ph]} \BibitemShut {NoStop}%
\bibitem [{\citenamefont {Johnson}\ \emph {et~al.}(2024)\citenamefont
  {Johnson}, \citenamefont {Polyakov}, \citenamefont {Skwarnicki},\ and\
  \citenamefont {Wang}}]{Johnson:2024omq}%
  \BibitemOpen
  \bibfield  {author} {\bibinfo {author} {\bibfnamefont {D.}~\bibnamefont
  {Johnson}}, \bibinfo {author} {\bibfnamefont {I.}~\bibnamefont {Polyakov}},
  \bibinfo {author} {\bibfnamefont {T.}~\bibnamefont {Skwarnicki}}, \ and\
  \bibinfo {author} {\bibfnamefont {M.}~\bibnamefont {Wang}},\ }\href {\doibase
  10.1146/annurev-nucl-102422-040628} {\  (\bibinfo {year} {2024}),\
  10.1146/annurev-nucl-102422-040628},\ \Eprint
  {http://arxiv.org/abs/2403.04051} {arXiv:2403.04051 [hep-ex]} \BibitemShut
  {NoStop}%
\bibitem [{\citenamefont {Entem}\ \emph {et~al.}(2025)\citenamefont {Entem},
  \citenamefont {Fern{\'a}ndez}, \citenamefont {Ortega},\ and\ \citenamefont
  {Segovia}}]{Entem:2025bqt}%
  \BibitemOpen
  \bibfield  {author} {\bibinfo {author} {\bibfnamefont {D.~R.}\ \bibnamefont
  {Entem}}, \bibinfo {author} {\bibfnamefont {F.}~\bibnamefont
  {Fern{\'a}ndez}}, \bibinfo {author} {\bibfnamefont {P.~G.}\ \bibnamefont
  {Ortega}}, \ and\ \bibinfo {author} {\bibfnamefont {J.}~\bibnamefont
  {Segovia}},\ }\href@noop {} {\  (\bibinfo {year} {2025})},\ \Eprint
  {http://arxiv.org/abs/2504.07897} {arXiv:2504.07897 [hep-ph]} \BibitemShut
  {NoStop}%
\bibitem [{\citenamefont {Hanhart}(2025)}]{Hanhart:2025bun}%
  \BibitemOpen
  \bibfield  {author} {\bibinfo {author} {\bibfnamefont {C.}~\bibnamefont
  {Hanhart}},\ }\href@noop {} {\  (\bibinfo {year} {2025})},\ \Eprint
  {http://arxiv.org/abs/2504.06043} {arXiv:2504.06043 [hep-ph]} \BibitemShut
  {NoStop}%
\bibitem [{\citenamefont {Wang}\ \emph {et~al.}(2025)\citenamefont {Wang},
  \citenamefont {Liu},\ and\ \citenamefont {Gao}}]{Wang:2025dur}%
  \BibitemOpen
  \bibfield  {author} {\bibinfo {author} {\bibfnamefont {X.}~\bibnamefont
  {Wang}}, \bibinfo {author} {\bibfnamefont {X.}~\bibnamefont {Liu}}, \ and\
  \bibinfo {author} {\bibfnamefont {Y.}~\bibnamefont {Gao}},\ }\href@noop {} {\
   (\bibinfo {year} {2025})},\ \Eprint {http://arxiv.org/abs/2502.15117}
  {arXiv:2502.15117 [hep-ex]} \BibitemShut {NoStop}%
\bibitem [{\citenamefont {Wang}(2025)}]{Wang:2025sic}%
  \BibitemOpen
  \bibfield  {author} {\bibinfo {author} {\bibfnamefont {Z.-G.}\ \bibnamefont
  {Wang}},\ }\href@noop {} {\  (\bibinfo {year} {2025})},\ \Eprint
  {http://arxiv.org/abs/2502.11351} {arXiv:2502.11351 [hep-ph]} \BibitemShut
  {NoStop}%
\bibitem [{\citenamefont {Francis}(2025)}]{Francis:2024fwf}%
  \BibitemOpen
  \bibfield  {author} {\bibinfo {author} {\bibfnamefont {A.}~\bibnamefont
  {Francis}},\ }\href {\doibase 10.1016/j.ppnp.2024.104143} {\bibfield
  {journal} {\bibinfo  {journal} {Prog. Part. Nucl. Phys.}\ }\textbf {\bibinfo
  {volume} {140}},\ \bibinfo {pages} {104143} (\bibinfo {year} {2025})},\
  \Eprint {http://arxiv.org/abs/2502.04701} {arXiv:2502.04701 [hep-lat]}
  \BibitemShut {NoStop}%
\bibitem [{\citenamefont {Chen}\ \emph {et~al.}(2025)\citenamefont {Chen},
  \citenamefont {Chen}, \citenamefont {Guo}, \citenamefont {Ma}, \citenamefont
  {Shen}, \citenamefont {Shou}, \citenamefont {Shou}, \citenamefont {Wang},
  \citenamefont {Wu},\ and\ \citenamefont {Zou}}]{Chen:2024eaq}%
  \BibitemOpen
  \bibfield  {author} {\bibinfo {author} {\bibfnamefont {J.-H.}\ \bibnamefont
  {Chen}}, \bibinfo {author} {\bibfnamefont {J.}~\bibnamefont {Chen}}, \bibinfo
  {author} {\bibfnamefont {F.-K.}\ \bibnamefont {Guo}}, \bibinfo {author}
  {\bibfnamefont {Y.-G.}\ \bibnamefont {Ma}}, \bibinfo {author} {\bibfnamefont
  {C.-P.}\ \bibnamefont {Shen}}, \bibinfo {author} {\bibfnamefont {Q.-Y.}\
  \bibnamefont {Shou}}, \bibinfo {author} {\bibfnamefont {Q.}~\bibnamefont
  {Shou}}, \bibinfo {author} {\bibfnamefont {Q.}~\bibnamefont {Wang}}, \bibinfo
  {author} {\bibfnamefont {J.-J.}\ \bibnamefont {Wu}}, \ and\ \bibinfo {author}
  {\bibfnamefont {B.-S.}\ \bibnamefont {Zou}},\ }\href {\doibase
  10.1007/s41365-025-01664-w} {\bibfield  {journal} {\bibinfo  {journal} {Nucl.
  Sci. Tech.}\ }\textbf {\bibinfo {volume} {36}},\ \bibinfo {pages} {55}
  (\bibinfo {year} {2025})},\ \Eprint {http://arxiv.org/abs/2411.18257}
  {arXiv:2411.18257 [hep-ph]} \BibitemShut {NoStop}%
\bibitem [{\citenamefont {H{\"u}sken}\ \emph {et~al.}(2025)\citenamefont
  {H{\"u}sken}, \citenamefont {Norella},\ and\ \citenamefont
  {Polyakov}}]{Husken:2024rdk}%
  \BibitemOpen
  \bibfield  {author} {\bibinfo {author} {\bibfnamefont {N.}~\bibnamefont
  {H{\"u}sken}}, \bibinfo {author} {\bibfnamefont {E.~S.}\ \bibnamefont
  {Norella}}, \ and\ \bibinfo {author} {\bibfnamefont {I.}~\bibnamefont
  {Polyakov}},\ }\href {\doibase 10.1142/S0217732325300022} {\bibfield
  {journal} {\bibinfo  {journal} {Mod. Phys. Lett. A}\ }\textbf {\bibinfo
  {volume} {40}},\ \bibinfo {pages} {2530002} (\bibinfo {year} {2025})},\
  \Eprint {http://arxiv.org/abs/2410.06923} {arXiv:2410.06923 [hep-ph]}
  \BibitemShut {NoStop}%
\bibitem [{\citenamefont {Jaffe}(1977)}]{Jaffe:1976yi}%
  \BibitemOpen
  \bibfield  {author} {\bibinfo {author} {\bibfnamefont {R.~L.}\ \bibnamefont
  {Jaffe}},\ }\href {\doibase 10.1103/PhysRevLett.38.195} {\bibfield  {journal}
  {\bibinfo  {journal} {Phys. Rev. Lett.}\ }\textbf {\bibinfo {volume} {38}},\
  \bibinfo {pages} {195} (\bibinfo {year} {1977})},\ \bibinfo {note} {[Erratum:
  Phys.Rev.Lett. 38, 617 (1977)]}\BibitemShut {NoStop}%
\bibitem [{\citenamefont {Mulders}\ \emph {et~al.}(1980)\citenamefont
  {Mulders}, \citenamefont {Aerts},\ and\ \citenamefont
  {De~Swart}}]{Mulders:1980vx}%
  \BibitemOpen
  \bibfield  {author} {\bibinfo {author} {\bibfnamefont {P.~J.}\ \bibnamefont
  {Mulders}}, \bibinfo {author} {\bibfnamefont {A.~T.~M.}\ \bibnamefont
  {Aerts}}, \ and\ \bibinfo {author} {\bibfnamefont {J.~J.}\ \bibnamefont
  {De~Swart}},\ }\href {\doibase 10.1103/PhysRevD.21.2653} {\bibfield
  {journal} {\bibinfo  {journal} {Phys. Rev. D}\ }\textbf {\bibinfo {volume}
  {21}},\ \bibinfo {pages} {2653} (\bibinfo {year} {1980})}\BibitemShut
  {NoStop}%
\bibitem [{\citenamefont {Kopeliovich}\ \emph {et~al.}(1990)\citenamefont
  {Kopeliovich}, \citenamefont {Schwesinger},\ and\ \citenamefont
  {Stern}}]{Kopeliovich:1990pp}%
  \BibitemOpen
  \bibfield  {author} {\bibinfo {author} {\bibfnamefont {V.~B.}\ \bibnamefont
  {Kopeliovich}}, \bibinfo {author} {\bibfnamefont {B.}~\bibnamefont
  {Schwesinger}}, \ and\ \bibinfo {author} {\bibfnamefont {B.~E.}\ \bibnamefont
  {Stern}},\ }\href {\doibase 10.1016/0370-2693(90)91451-G} {\bibfield
  {journal} {\bibinfo  {journal} {Phys. Lett. B}\ }\textbf {\bibinfo {volume}
  {242}},\ \bibinfo {pages} {145} (\bibinfo {year} {1990})}\BibitemShut
  {NoStop}%
\bibitem [{\citenamefont {Leandri}\ and\ \citenamefont
  {Silvestre-Brac}(1995)}]{Leandri:1995zm}%
  \BibitemOpen
  \bibfield  {author} {\bibinfo {author} {\bibfnamefont {J.}~\bibnamefont
  {Leandri}}\ and\ \bibinfo {author} {\bibfnamefont {B.}~\bibnamefont
  {Silvestre-Brac}},\ }\href {\doibase 10.1103/PhysRevD.51.3628} {\bibfield
  {journal} {\bibinfo  {journal} {Phys. Rev. D}\ }\textbf {\bibinfo {volume}
  {51}},\ \bibinfo {pages} {3628} (\bibinfo {year} {1995})}\BibitemShut
  {NoStop}%
\bibitem [{\citenamefont {Pepin}\ and\ \citenamefont
  {Stancu}(1998)}]{Pepin:1998ih}%
  \BibitemOpen
  \bibfield  {author} {\bibinfo {author} {\bibfnamefont {S.}~\bibnamefont
  {Pepin}}\ and\ \bibinfo {author} {\bibfnamefont {F.}~\bibnamefont {Stancu}},\
  }\href {\doibase 10.1103/PhysRevD.57.4475} {\bibfield  {journal} {\bibinfo
  {journal} {Phys. Rev. D}\ }\textbf {\bibinfo {volume} {57}},\ \bibinfo
  {pages} {4475} (\bibinfo {year} {1998})},\ \Eprint
  {http://arxiv.org/abs/hep-ph/9710528} {arXiv:hep-ph/9710528} \BibitemShut
  {NoStop}%
\bibitem [{\citenamefont {Li}\ \emph {et~al.}(2001)\citenamefont {Li},
  \citenamefont {Shen}, \citenamefont {Zhang},\ and\ \citenamefont
  {Yu}}]{Li:2000cb}%
  \BibitemOpen
  \bibfield  {author} {\bibinfo {author} {\bibfnamefont {Q.~B.}\ \bibnamefont
  {Li}}, \bibinfo {author} {\bibfnamefont {P.~N.}\ \bibnamefont {Shen}},
  \bibinfo {author} {\bibfnamefont {Z.~Y.}\ \bibnamefont {Zhang}}, \ and\
  \bibinfo {author} {\bibfnamefont {Y.~W.}\ \bibnamefont {Yu}},\ }\href
  {\doibase 10.1016/S0375-9474(00)00458-9} {\bibfield  {journal} {\bibinfo
  {journal} {Nucl. Phys. A}\ }\textbf {\bibinfo {volume} {683}},\ \bibinfo
  {pages} {487} (\bibinfo {year} {2001})},\ \Eprint
  {http://arxiv.org/abs/nucl-th/0009038} {arXiv:nucl-th/0009038} \BibitemShut
  {NoStop}%
\bibitem [{\citenamefont {Herzog}\ and\ \citenamefont
  {McKernan}(2003)}]{Herzog:2003wt}%
  \BibitemOpen
  \bibfield  {author} {\bibinfo {author} {\bibfnamefont {C.~P.}\ \bibnamefont
  {Herzog}}\ and\ \bibinfo {author} {\bibfnamefont {J.}~\bibnamefont
  {McKernan}},\ }\href {\doibase 10.1088/1126-6708/2003/08/054} {\bibfield
  {journal} {\bibinfo  {journal} {JHEP}\ }\textbf {\bibinfo {volume} {08}},\
  \bibinfo {pages} {054} (\bibinfo {year} {2003})},\ \Eprint
  {http://arxiv.org/abs/hep-th/0305048} {arXiv:hep-th/0305048} \BibitemShut
  {NoStop}%
\bibitem [{\citenamefont {Valcarce}\ \emph {et~al.}(2005)\citenamefont
  {Valcarce}, \citenamefont {Garcilazo}, \citenamefont {Fernandez},\ and\
  \citenamefont {Gonzalez}}]{Valcarce:2005em}%
  \BibitemOpen
  \bibfield  {author} {\bibinfo {author} {\bibfnamefont {A.}~\bibnamefont
  {Valcarce}}, \bibinfo {author} {\bibfnamefont {H.}~\bibnamefont {Garcilazo}},
  \bibinfo {author} {\bibfnamefont {F.}~\bibnamefont {Fernandez}}, \ and\
  \bibinfo {author} {\bibfnamefont {P.}~\bibnamefont {Gonzalez}},\ }\href
  {\doibase 10.1088/0034-4885/68/5/R01} {\bibfield  {journal} {\bibinfo
  {journal} {Rept. Prog. Phys.}\ }\textbf {\bibinfo {volume} {68}},\ \bibinfo
  {pages} {965} (\bibinfo {year} {2005})},\ \Eprint
  {http://arxiv.org/abs/hep-ph/0502173} {arXiv:hep-ph/0502173} \BibitemShut
  {NoStop}%
\bibitem [{\citenamefont {Ikeda}\ and\ \citenamefont
  {Sato}(2007)}]{Ikeda:2007nz}%
  \BibitemOpen
  \bibfield  {author} {\bibinfo {author} {\bibfnamefont {Y.}~\bibnamefont
  {Ikeda}}\ and\ \bibinfo {author} {\bibfnamefont {T.}~\bibnamefont {Sato}},\
  }\href {\doibase 10.1103/PhysRevC.76.035203} {\bibfield  {journal} {\bibinfo
  {journal} {Phys. Rev. C}\ }\textbf {\bibinfo {volume} {76}},\ \bibinfo
  {pages} {035203} (\bibinfo {year} {2007})},\ \Eprint
  {http://arxiv.org/abs/0704.1978} {arXiv:0704.1978 [nucl-th]} \BibitemShut
  {NoStop}%
\bibitem [{\citenamefont {Olsen}(2015)}]{Olsen:2014qna}%
  \BibitemOpen
  \bibfield  {author} {\bibinfo {author} {\bibfnamefont {S.~L.}\ \bibnamefont
  {Olsen}},\ }\href {\doibase 10.1007/S11467-014-0449-6} {\bibfield  {journal}
  {\bibinfo  {journal} {Front. Phys. (Beijing)}\ }\textbf {\bibinfo {volume}
  {10}},\ \bibinfo {pages} {121} (\bibinfo {year} {2015})},\ \Eprint
  {http://arxiv.org/abs/1411.7738} {arXiv:1411.7738 [hep-ex]} \BibitemShut
  {NoStop}%
\bibitem [{\citenamefont {Francis}\ \emph {et~al.}(2019)\citenamefont
  {Francis}, \citenamefont {Green}, \citenamefont {Junnarkar}, \citenamefont
  {Miao}, \citenamefont {Rae},\ and\ \citenamefont {Wittig}}]{Francis:2018qch}%
  \BibitemOpen
  \bibfield  {author} {\bibinfo {author} {\bibfnamefont {A.}~\bibnamefont
  {Francis}}, \bibinfo {author} {\bibfnamefont {J.~R.}\ \bibnamefont {Green}},
  \bibinfo {author} {\bibfnamefont {P.~M.}\ \bibnamefont {Junnarkar}}, \bibinfo
  {author} {\bibfnamefont {C.}~\bibnamefont {Miao}}, \bibinfo {author}
  {\bibfnamefont {T.~D.}\ \bibnamefont {Rae}}, \ and\ \bibinfo {author}
  {\bibfnamefont {H.}~\bibnamefont {Wittig}},\ }\href {\doibase
  10.1103/PhysRevD.99.074505} {\bibfield  {journal} {\bibinfo  {journal} {Phys.
  Rev. D}\ }\textbf {\bibinfo {volume} {99}},\ \bibinfo {pages} {074505}
  (\bibinfo {year} {2019})},\ \Eprint {http://arxiv.org/abs/1805.03966}
  {arXiv:1805.03966 [hep-lat]} \BibitemShut {NoStop}%
\bibitem [{\citenamefont {Amarasinghe}\ \emph {et~al.}(2023)\citenamefont
  {Amarasinghe}, \citenamefont {Baghdadi}, \citenamefont {Davoudi},
  \citenamefont {Detmold}, \citenamefont {Illa}, \citenamefont {Parreno},
  \citenamefont {Pochinsky}, \citenamefont {Shanahan},\ and\ \citenamefont
  {Wagman}}]{Amarasinghe:2021lqa}%
  \BibitemOpen
  \bibfield  {author} {\bibinfo {author} {\bibfnamefont {S.}~\bibnamefont
  {Amarasinghe}}, \bibinfo {author} {\bibfnamefont {R.}~\bibnamefont
  {Baghdadi}}, \bibinfo {author} {\bibfnamefont {Z.}~\bibnamefont {Davoudi}},
  \bibinfo {author} {\bibfnamefont {W.}~\bibnamefont {Detmold}}, \bibinfo
  {author} {\bibfnamefont {M.}~\bibnamefont {Illa}}, \bibinfo {author}
  {\bibfnamefont {A.}~\bibnamefont {Parreno}}, \bibinfo {author} {\bibfnamefont
  {A.~V.}\ \bibnamefont {Pochinsky}}, \bibinfo {author} {\bibfnamefont {P.~E.}\
  \bibnamefont {Shanahan}}, \ and\ \bibinfo {author} {\bibfnamefont {M.~L.}\
  \bibnamefont {Wagman}},\ }\href {\doibase 10.1103/PhysRevD.107.094508}
  {\bibfield  {journal} {\bibinfo  {journal} {Phys. Rev. D}\ }\textbf {\bibinfo
  {volume} {107}},\ \bibinfo {pages} {094508} (\bibinfo {year} {2023})},\
  \bibinfo {note} {[Erratum: Phys.Rev.D 110, 119904 (2024)]},\ \Eprint
  {http://arxiv.org/abs/2108.10835} {arXiv:2108.10835 [hep-lat]} \BibitemShut
  {NoStop}%
\bibitem [{\citenamefont {Wan}\ \emph {et~al.}(2022)\citenamefont {Wan},
  \citenamefont {Zhang},\ and\ \citenamefont {Qiao}}]{Wan:2021vny}%
  \BibitemOpen
  \bibfield  {author} {\bibinfo {author} {\bibfnamefont {B.-D.}\ \bibnamefont
  {Wan}}, \bibinfo {author} {\bibfnamefont {S.-Q.}\ \bibnamefont {Zhang}}, \
  and\ \bibinfo {author} {\bibfnamefont {C.-F.}\ \bibnamefont {Qiao}},\ }\href
  {\doibase 10.1103/PhysRevD.105.014016} {\bibfield  {journal} {\bibinfo
  {journal} {Phys. Rev. D}\ }\textbf {\bibinfo {volume} {105}},\ \bibinfo
  {pages} {014016} (\bibinfo {year} {2022})},\ \Eprint
  {http://arxiv.org/abs/2109.07130} {arXiv:2109.07130 [hep-ph]} \BibitemShut
  {NoStop}%
\bibitem [{\citenamefont {Wang}\ \emph {et~al.}(2024)\citenamefont {Wang},
  \citenamefont {Chen}, \citenamefont {Meng},\ and\ \citenamefont
  {Zhu}}]{Wang:2024riu}%
  \BibitemOpen
  \bibfield  {author} {\bibinfo {author} {\bibfnamefont {B.}~\bibnamefont
  {Wang}}, \bibinfo {author} {\bibfnamefont {K.}~\bibnamefont {Chen}}, \bibinfo
  {author} {\bibfnamefont {L.}~\bibnamefont {Meng}}, \ and\ \bibinfo {author}
  {\bibfnamefont {S.-L.}\ \bibnamefont {Zhu}},\ }\href {\doibase
  10.1103/PhysRevD.110.014038} {\bibfield  {journal} {\bibinfo  {journal}
  {Phys. Rev. D}\ }\textbf {\bibinfo {volume} {110}},\ \bibinfo {pages}
  {014038} (\bibinfo {year} {2024})},\ \Eprint
  {http://arxiv.org/abs/2406.06993} {arXiv:2406.06993 [hep-ph]} \BibitemShut
  {NoStop}%
\bibitem [{\citenamefont {Hammond}\ \emph {et~al.}(1994)\citenamefont
  {Hammond}, \citenamefont {Lester},\ and\ \citenamefont
  {Reynolds}}]{Hammond:1994bk}%
  \BibitemOpen
  \bibfield  {author} {\bibinfo {author} {\bibfnamefont {B.}~\bibnamefont
  {Hammond}}, \bibinfo {author} {\bibfnamefont {W.}~\bibnamefont {Lester}}, \
  and\ \bibinfo {author} {\bibfnamefont {P.}~\bibnamefont {Reynolds}},\
  }\href@noop {} {\emph {\bibinfo {title} {Monte Carlo Methods in ab Initio
  Quantum Chemistry}}}\ (\bibinfo  {publisher} {World Scientific},\ \bibinfo
  {address} {Singapore},\ \bibinfo {year} {1994})\BibitemShut {NoStop}%
\bibitem [{\citenamefont {Foulkes}\ \emph {et~al.}(2001)\citenamefont
  {Foulkes}, \citenamefont {Mitas}, \citenamefont {Needs},\ and\ \citenamefont
  {Rajagopal}}]{Foulkes:2001zz}%
  \BibitemOpen
  \bibfield  {author} {\bibinfo {author} {\bibfnamefont {W.}~\bibnamefont
  {Foulkes}}, \bibinfo {author} {\bibfnamefont {L.}~\bibnamefont {Mitas}},
  \bibinfo {author} {\bibfnamefont {R.}~\bibnamefont {Needs}}, \ and\ \bibinfo
  {author} {\bibfnamefont {G.}~\bibnamefont {Rajagopal}},\ }\href {\doibase
  10.1103/RevModPhys.73.33} {\bibfield  {journal} {\bibinfo  {journal} {Rev.
  Mod. Phys.}\ }\textbf {\bibinfo {volume} {73}},\ \bibinfo {pages} {33}
  (\bibinfo {year} {2001})}\BibitemShut {NoStop}%
\bibitem [{\citenamefont {Nightingale}\ and\ \citenamefont
  {Umrigar}(2014)}]{Nightingale:2014bk}%
  \BibitemOpen
  \bibfield  {author} {\bibinfo {author} {\bibfnamefont {M.}~\bibnamefont
  {Nightingale}}\ and\ \bibinfo {author} {\bibfnamefont {C.~J.}\ \bibnamefont
  {Umrigar}},\ }\href@noop {} {\emph {\bibinfo {title} {Quantum Monte Carlo
  Methods in Physics and Chemistry}}}\ (\bibinfo  {publisher} {Springer},\
  \bibinfo {address} {Vienna},\ \bibinfo {year} {2014})\BibitemShut {NoStop}%
\bibitem [{\citenamefont {Ceperley}(1995)}]{Ceperley:1995zz}%
  \BibitemOpen
  \bibfield  {author} {\bibinfo {author} {\bibfnamefont {D.}~\bibnamefont
  {Ceperley}},\ }\href {\doibase 10.1103/RevModPhys.67.279} {\bibfield
  {journal} {\bibinfo  {journal} {Rev. Mod. Phys.}\ }\textbf {\bibinfo {volume}
  {67}},\ \bibinfo {pages} {279} (\bibinfo {year} {1995})}\BibitemShut
  {NoStop}%
\bibitem [{\citenamefont {Giorgini}\ \emph {et~al.}(2008)\citenamefont
  {Giorgini}, \citenamefont {Pitaevskii},\ and\ \citenamefont
  {Stringari}}]{Giorgini:2008zz}%
  \BibitemOpen
  \bibfield  {author} {\bibinfo {author} {\bibfnamefont {S.}~\bibnamefont
  {Giorgini}}, \bibinfo {author} {\bibfnamefont {L.~P.}\ \bibnamefont
  {Pitaevskii}}, \ and\ \bibinfo {author} {\bibfnamefont {S.}~\bibnamefont
  {Stringari}},\ }\href {\doibase 10.1103/RevModPhys.80.1215} {\bibfield
  {journal} {\bibinfo  {journal} {Rev. Mod. Phys.}\ }\textbf {\bibinfo {volume}
  {80}},\ \bibinfo {pages} {1215} (\bibinfo {year} {2008})},\ \Eprint
  {http://arxiv.org/abs/0706.3360} {arXiv:0706.3360 [cond-mat.other]}
  \BibitemShut {NoStop}%
\bibitem [{\citenamefont {Carlson}\ \emph
  {et~al.}(1983{\natexlab{a}})\citenamefont {Carlson}, \citenamefont {Kogut},\
  and\ \citenamefont {Pandharipande}}]{Carlson:1982xi}%
  \BibitemOpen
  \bibfield  {author} {\bibinfo {author} {\bibfnamefont {J.}~\bibnamefont
  {Carlson}}, \bibinfo {author} {\bibfnamefont {J.~B.}\ \bibnamefont {Kogut}},
  \ and\ \bibinfo {author} {\bibfnamefont {V.}~\bibnamefont {Pandharipande}},\
  }\href {\doibase 10.1103/PhysRevD.27.233} {\bibfield  {journal} {\bibinfo
  {journal} {Phys. Rev. D}\ }\textbf {\bibinfo {volume} {27}},\ \bibinfo
  {pages} {233} (\bibinfo {year} {1983}{\natexlab{a}})}\BibitemShut {NoStop}%
\bibitem [{\citenamefont {Carlson}\ \emph
  {et~al.}(1983{\natexlab{b}})\citenamefont {Carlson}, \citenamefont {Kogut},\
  and\ \citenamefont {Pandharipande}}]{Carlson:1983rw}%
  \BibitemOpen
  \bibfield  {author} {\bibinfo {author} {\bibfnamefont {J.}~\bibnamefont
  {Carlson}}, \bibinfo {author} {\bibfnamefont {J.}~\bibnamefont {Kogut}}, \
  and\ \bibinfo {author} {\bibfnamefont {V.}~\bibnamefont {Pandharipande}},\
  }\href {\doibase 10.1103/PhysRevD.28.2807} {\bibfield  {journal} {\bibinfo
  {journal} {Phys. Rev. D}\ }\textbf {\bibinfo {volume} {28}},\ \bibinfo
  {pages} {2807} (\bibinfo {year} {1983}{\natexlab{b}})}\BibitemShut {NoStop}%
\bibitem [{\citenamefont {Bai}\ \emph {et~al.}(2019)\citenamefont {Bai},
  \citenamefont {Lu},\ and\ \citenamefont {Osborne}}]{Bai:2016int}%
  \BibitemOpen
  \bibfield  {author} {\bibinfo {author} {\bibfnamefont {Y.}~\bibnamefont
  {Bai}}, \bibinfo {author} {\bibfnamefont {S.}~\bibnamefont {Lu}}, \ and\
  \bibinfo {author} {\bibfnamefont {J.}~\bibnamefont {Osborne}},\ }\href
  {\doibase 10.1016/j.physletb.2019.134930} {\bibfield  {journal} {\bibinfo
  {journal} {Phys. Lett. B}\ }\textbf {\bibinfo {volume} {798}},\ \bibinfo
  {pages} {134930} (\bibinfo {year} {2019})},\ \Eprint
  {http://arxiv.org/abs/1612.00012} {arXiv:1612.00012 [hep-ph]} \BibitemShut
  {NoStop}%
\bibitem [{\citenamefont {Gordillo}\ \emph {et~al.}(2020)\citenamefont
  {Gordillo}, \citenamefont {De~Soto},\ and\ \citenamefont
  {Segovia}}]{Gordillo:2020sgc}%
  \BibitemOpen
  \bibfield  {author} {\bibinfo {author} {\bibfnamefont {M.~C.}\ \bibnamefont
  {Gordillo}}, \bibinfo {author} {\bibfnamefont {F.}~\bibnamefont {De~Soto}}, \
  and\ \bibinfo {author} {\bibfnamefont {J.}~\bibnamefont {Segovia}},\ }\href
  {\doibase 10.1103/PhysRevD.102.114007} {\bibfield  {journal} {\bibinfo
  {journal} {Phys. Rev. D}\ }\textbf {\bibinfo {volume} {102}},\ \bibinfo
  {pages} {114007} (\bibinfo {year} {2020})},\ \Eprint
  {http://arxiv.org/abs/2009.11889} {arXiv:2009.11889 [hep-ph]} \BibitemShut
  {NoStop}%
\bibitem [{\citenamefont {Semay}\ and\ \citenamefont
  {Silvestre-Brac}(1994)}]{Semay:1994ht}%
  \BibitemOpen
  \bibfield  {author} {\bibinfo {author} {\bibfnamefont {C.}~\bibnamefont
  {Semay}}\ and\ \bibinfo {author} {\bibfnamefont {B.}~\bibnamefont
  {Silvestre-Brac}},\ }\href {\doibase 10.1007/BF01413104} {\bibfield
  {journal} {\bibinfo  {journal} {Z. Phys. C}\ }\textbf {\bibinfo {volume}
  {61}},\ \bibinfo {pages} {271} (\bibinfo {year} {1994})}\BibitemShut
  {NoStop}%
\bibitem [{\citenamefont {Silvestre-Brac}(1996)}]{Silvestre-Brac:1996myf}%
  \BibitemOpen
  \bibfield  {author} {\bibinfo {author} {\bibfnamefont {B.}~\bibnamefont
  {Silvestre-Brac}},\ }\href {\doibase 10.1007/s006010050028} {\bibfield
  {journal} {\bibinfo  {journal} {Few Body Syst.}\ }\textbf {\bibinfo {volume}
  {20}},\ \bibinfo {pages} {1} (\bibinfo {year} {1996})}\BibitemShut {NoStop}%
\bibitem [{\citenamefont {Gordillo}(2026)}]{Gordillo:2026usv}%
  \BibitemOpen
  \bibfield  {author} {\bibinfo {author} {\bibfnamefont {M.~C.}\ \bibnamefont
  {Gordillo}},\ }\href@noop {} {\  (\bibinfo {year} {2026})},\ \Eprint
  {http://arxiv.org/abs/2604.18174} {arXiv:2604.18174 [hep-ph]} \BibitemShut
  {NoStop}%
\bibitem [{\citenamefont {Tong}()}]{TongTopicsQM}%
  \BibitemOpen
  \bibfield  {author} {\bibinfo {author} {\bibfnamefont {D.}~\bibnamefont
  {Tong}},\ }\href {https://www.damtp.cam.ac.uk/user/tong/topicsinqm.html}
  {\enquote {\bibinfo {title} {Topics in quantum mechanics},}\ }\bibinfo
  {howpublished} {Lecture notes, University of Cambridge},\ \bibinfo {note}
  {sec. 5.4.2, ``Generalised Measurements'', pp. 176--177}\BibitemShut
  {NoStop}%
\bibitem [{\citenamefont {Gordillo}\ and\ \citenamefont
  {Alcaraz-Pelegrina}(2023)}]{Gordillo:2023tnz}%
  \BibitemOpen
  \bibfield  {author} {\bibinfo {author} {\bibfnamefont {M.~C.}\ \bibnamefont
  {Gordillo}}\ and\ \bibinfo {author} {\bibfnamefont {J.~M.}\ \bibnamefont
  {Alcaraz-Pelegrina}},\ }\href {\doibase 10.1103/PhysRevD.108.054027}
  {\bibfield  {journal} {\bibinfo  {journal} {Phys. Rev. D}\ }\textbf {\bibinfo
  {volume} {108}},\ \bibinfo {pages} {054027} (\bibinfo {year} {2023})},\
  \Eprint {http://arxiv.org/abs/2307.08408} {arXiv:2307.08408 [hep-ph]}
  \BibitemShut {NoStop}%
\bibitem [{\citenamefont {Dyson}\ and\ \citenamefont {Xuong}(1964)}]{dyson}%
  \BibitemOpen
  \bibfield  {author} {\bibinfo {author} {\bibfnamefont {F.~J.}\ \bibnamefont
  {Dyson}}\ and\ \bibinfo {author} {\bibfnamefont {N.-H.}\ \bibnamefont
  {Xuong}},\ }\href {\doibase 10.1103/PhysRevLett.13.815} {\bibfield  {journal}
  {\bibinfo  {journal} {Phys. Rev. Lett.}\ }\textbf {\bibinfo {volume} {13}},\
  \bibinfo {pages} {815} (\bibinfo {year} {1964})}\BibitemShut {NoStop}%
\bibitem [{\citenamefont {Gordillo}\ \emph {et~al.}(2021)\citenamefont
  {Gordillo}, \citenamefont {De~Soto},\ and\ \citenamefont
  {Segovia}}]{Gordillo:2021bra}%
  \BibitemOpen
  \bibfield  {author} {\bibinfo {author} {\bibfnamefont {M.~C.}\ \bibnamefont
  {Gordillo}}, \bibinfo {author} {\bibfnamefont {F.}~\bibnamefont {De~Soto}}, \
  and\ \bibinfo {author} {\bibfnamefont {J.}~\bibnamefont {Segovia}},\ }\href
  {\doibase 10.1103/PhysRevD.104.054036} {\bibfield  {journal} {\bibinfo
  {journal} {Phys. Rev. D}\ }\textbf {\bibinfo {volume} {104}},\ \bibinfo
  {pages} {054036} (\bibinfo {year} {2021})},\ \Eprint
  {http://arxiv.org/abs/2105.11976} {arXiv:2105.11976 [hep-ph]} \BibitemShut
  {NoStop}%
\bibitem [{\citenamefont {Gordillo}\ \emph {et~al.}(2024)\citenamefont
  {Gordillo}, \citenamefont {Segovia},\ and\ \citenamefont
  {Alcaraz-Pelegrina}}]{Gordillo:2024blx}%
  \BibitemOpen
  \bibfield  {author} {\bibinfo {author} {\bibfnamefont {M.~C.}\ \bibnamefont
  {Gordillo}}, \bibinfo {author} {\bibfnamefont {J.}~\bibnamefont {Segovia}}, \
  and\ \bibinfo {author} {\bibfnamefont {J.~M.}\ \bibnamefont
  {Alcaraz-Pelegrina}},\ }\href {\doibase 10.1103/PhysRevD.110.094024}
  {\bibfield  {journal} {\bibinfo  {journal} {Phys. Rev. D}\ }\textbf {\bibinfo
  {volume} {110}},\ \bibinfo {pages} {094024} (\bibinfo {year} {2024})},\
  \Eprint {http://arxiv.org/abs/2409.04130} {arXiv:2409.04130 [hep-ph]}
  \BibitemShut {NoStop}%
\bibitem [{\citenamefont {Silvestre-Brac}\ and\ \citenamefont
  {Leandri}(1992)}]{Silvestre-Brac:1992xsl}%
  \BibitemOpen
  \bibfield  {author} {\bibinfo {author} {\bibfnamefont {B.}~\bibnamefont
  {Silvestre-Brac}}\ and\ \bibinfo {author} {\bibfnamefont {J.}~\bibnamefont
  {Leandri}},\ }\href {\doibase 10.1103/PhysRevD.45.4221} {\bibfield  {journal}
  {\bibinfo  {journal} {Phys. Rev. D}\ }\textbf {\bibinfo {volume} {45}},\
  \bibinfo {pages} {4221} (\bibinfo {year} {1992})}\BibitemShut {NoStop}%
\bibitem [{\citenamefont {Kessler}\ \emph {et~al.}(1999)\citenamefont
  {Kessler}, \citenamefont {Dewey}, \citenamefont {Deslattes}, \citenamefont
  {Henins}, \citenamefont {B{\"o}rner}, \citenamefont {Jentschel},
  \citenamefont {Doll},\ and\ \citenamefont {Lehmann}}]{Kessler1999}%
  \BibitemOpen
  \bibfield  {author} {\bibinfo {author} {\bibfnamefont {E.~G.~J.}\
  \bibnamefont {Kessler}}, \bibinfo {author} {\bibfnamefont {M.~S.}\
  \bibnamefont {Dewey}}, \bibinfo {author} {\bibfnamefont {R.~D.}\ \bibnamefont
  {Deslattes}}, \bibinfo {author} {\bibfnamefont {A.}~\bibnamefont {Henins}},
  \bibinfo {author} {\bibfnamefont {H.~G.}\ \bibnamefont {B{\"o}rner}},
  \bibinfo {author} {\bibfnamefont {M.}~\bibnamefont {Jentschel}}, \bibinfo
  {author} {\bibfnamefont {C.}~\bibnamefont {Doll}}, \ and\ \bibinfo {author}
  {\bibfnamefont {H.}~\bibnamefont {Lehmann}},\ }\href {\doibase
  10.1016/S0375-9601(99)00078-X} {\bibfield  {journal} {\bibinfo  {journal}
  {Physics Letters A}\ }\textbf {\bibinfo {volume} {255}},\ \bibinfo {pages}
  {221} (\bibinfo {year} {1999})}\BibitemShut {NoStop}%
\end{thebibliography}%

\end{document}